\documentclass[12pt]{article}
\usepackage{epsfig}

\def\be{\begin{eqnarray}}
\def\ee{\end{eqnarray}}
\def\bq{\begin{equation}}
\def\eq{\end{equation}}
\def\ben{\begin{enumerate}}
\def\een{\end{enumerate}}

\def\roughly#1{\mathrel{\raise.3ex\hbox{$#1$\kern-.75em%
\lower1ex\hbox{$\sim$}}}}

\newcommand{\nslash}{\kern 0.2 em n\kern -0.50em /}

\newcommand{\beq}{\begin{eqnarray}}
\newcommand{\eeq}{\end{eqnarray}}

\begin{document}
\begin{titlepage}

 \vspace{1.5cm}
\begin{center}
\ \\
{\bf\large Helicity dependent parton distributions}

\vskip 0.7cm

{\bf Sergio Scopetta}

{\it Dipartimento di Fisica, Universit\`a degli Studi di Perugia and INFN, sezione di Perugia, via A. Pascoli
06100 Perugia, Italy.}

{\small Email: sergio.scopetta@pg.infn.it}

\vspace{0.7cm}

{\bf Vicente Vento}

{\it  Departamento de F\'{\i}sica Te\'orica and Instituto de
F\'{\i}sica Corpuscular, Universidad de Valencia - Consejo Superior de Investigaciones
Cient\'{\i}ficas, 46100 Burjassot (Val\`encia), Spain. }

{\small Email: vicente.vento@uv.es}

\end{center}

The helicity dependent parton distributions describe the number 
density of partons with given longitudinal momentum $x$ and given 
polarization in a hadron polarized longitudinally with respect to its motion. 
After the discovery, more than 70 years ago, that the proton is not 
elementary, the observation of Bjorken scaling in the late 1960s lead to 
the idea of hadrons containing almost pointlike constituents, the partons. 
Since then, Deep Inelastic Scattering (DIS) has played a crucial role in our 
understanding of hadron structure. Through DIS experiments it has been 
possible to link the partons to the quarks, and to unveil the presence of 
other pointlike constituents, the gluons, which lead into a dynamical theory 
of quarks and gluons - quantum chromodynamics (QCD). 
Polarized DIS, i.e. the collision of a longitudinally polarized lepton 
beam on a polarized target (either longitudinally or transversely polarized), 
provides a complementary information regarding the structure of the nucleon. 
Whereas ordinary DIS probes simply the number density of partons with a 
fraction $x$ of the momentum of the parent hadron, polarized DIS can 
partly answer the question as to the number density of partons with given $x$ 
and given spin polarization in a hadron of definite polarization, 
either parallel or transverse with respect to the motion of the hadron. 
In here, the phenomenology associated with DIS off longitudinally polarized 
targets, and the present theoretical understanding of the dynamics involved, 
are described.

\end{titlepage}

{\bf Index}

1. Introduction

2. Spin dependent structure functions

3. The parton model and QCD

4. The proton spin problem

5. Models of hadron structure

6. Experimental determination of the longitudinal polarization functions

7. Summary

8. References

\section{Introduction}

Understanding the spin structure of the proton is one of the most challenging 
problems in nowadays subatomic
physics. The story of the proton spin dates back to the  measurement, 
in 1933, of its anomalous
magnetic moment, $\kappa_p \simeq 1.79$ Bohr magnetons, revealing that the 
proton has an internal
structure. We now understand the proton as a bound state of three 
confined valence quarks of spin 1/2,
interacting in a complicated manner through spin-1 colored gluons. 
How is the proton built up from its
constituents, quarks and gluons, and how is its spin distributed 
among them, are the questions that
polarized DIS experiments and their theoretical interpretation aim to answer.
DIS has always played a crucial role in our understanding of
the structure of hadrons. The observation of Bjorken scaling 
(Bjorken and Paschos, 1969 \cite{uno}) in the 
DIS experiments in the late 1960s  (Breidenbach {\it et al.}, 1969 \cite{due}) 
lead
to the idea that hadrons contain almost point like constituents, the partons.
Later, through DIS experiments, it was possible to link
the partons to the quarks, and to discover the existence of 
electrically neutral
constituents, the gluons, which lead into a dynamical theory of 
quarks and gluons - quantum
chromodynamics (QCD)(Fritzsch {\it et al.}, 1973 \cite{tre}; 
see also H.D. Politzer, 1974 \cite{quattro}). 
As a matter of facts, the study of the evolution with the momentum transfer
of DIS observables represents
probably the most direct test of the perturbative aspects of QCD.

Polarized DIS, involving the collision of a longitudinally polarized 
lepton beam on a polarized target
(either longitudinally or transversely polarized) provides a 
complementary and important insight into
the structure of the nucleon. Whereas ordinary DIS probes 
simply the number density of partons with a
fraction $x$ of the momentum of the parent hadron, polarized DIS can 
partly answer the more sophisticated
question as to the number density of partons with given $x$ and 
given helicity, in a hadron of
definite polarization, either parallel or transverse to its motion.
The polarized DIS cross section is described by two spin dependent 
structure functions, 
$g_1$ and $g_2$. The study of the longitudinal polarization structure 
function $g_1$
for a long time remained comfortably at the level of partons. In 1988,
the proton data
of the European Muon Collaboration (EMC)(Ashman {\it et al.}, 1988 
\cite{cinque}), 
which differed significantly from naive theoretical
predictions, were published. Those results were argued to 
imply that the sum of the spins carried by the quarks in
a proton  was consistent with zero, rather than with 1/2,  
the non-relativistic quark model value, suggesting a
spin crisis in the parton model. 
In the following, the phenomenology associated with
DIS off longitudinally polarized targets, and the present theoretical 
understanding of the dynamics involved, are described.
At the end, the interested reader will find an extended bibliography 
of up to date professional reviews, and a series of links to web pages, 
where he can satisfy his quest for knowledge and detail, 
and find the original references.

\section{Polarized Parton Distributions}

High-energy lepton scattering off the nucleon,
i.e. the process $\ell N \rightarrow \ell^\prime X$, illustrated in Fig.1,
is called Deep ($-q^2=Q^2 >> M^2$) Inelastic ($W^2 >> M^2$) Scattering (DIS).
The filled circle in this figure represents the internal structure 
of the nucleon
which can be expressed in terms of structure functions.

\begin{figure}
\centerline{\epsfig{file=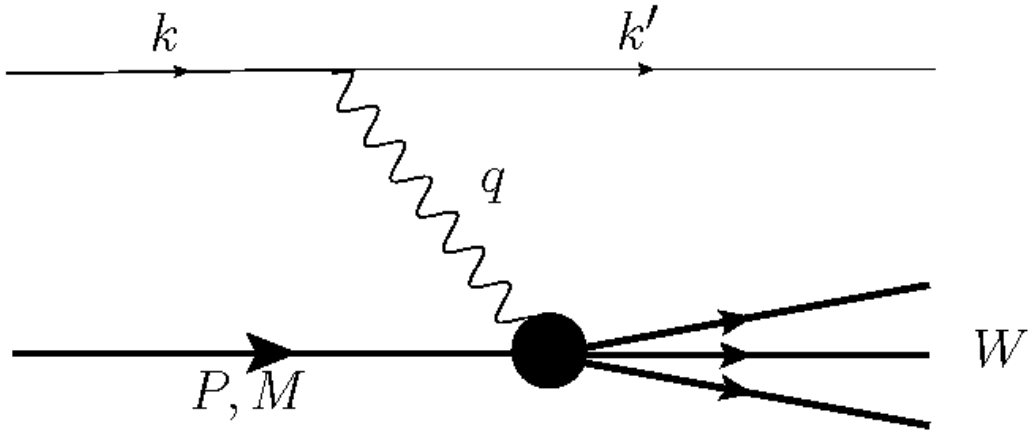,scale=0.5}}
\vskip 0.4cm
\caption{Kinematics of the DIS process.
The quantities $k =(E,\vec k)$ and $k^\prime=(E',\vec k')$
are the four-momenta of the incoming 
and outgoing leptons,
$P$ is the four-momentum of a nucleon with mass $M$ and $W$ is the mass of 
the recoiling system $X$.
We discuss here only
the exchange of a virtual photon.
The four-momentum transferred to the nucleon is
$q = k - k^\prime$.}
\end{figure}

The cross-section for inclusive unpolarized scattering can be written
 \begin{equation}
\label{eq:UnpolXsec}
 \frac{d^2 \sigma }{dx\,dy}=\frac{4 \pi  \alpha^2
 }{x\, y\, Q^2} \left\{  xy^2 F_1(x,Q^2) +
 \left[1-y-\frac{M^2x^2 y^2}{Q^2}\right] F_2(x,Q^2) \right\},
 \end{equation}
where $F_1$ and $F_2$ are the so called unpolarized structure functions.
The following notation has been used,
\begin{equation} \label{eq:def-Q,xBj}
%  Q^2 \equiv -q^2 \quad , \quad
x \equiv {Q^2 \over 2q \cdot P}= {Q^2 \over 2M\nu} \quad , \quad 
y\equiv 
\frac{P\cdot q}{P\cdot k}
=
\frac{\nu}{E} 
\end{equation}
where $\nu=E-E'$ is the energy of the virtual photon in the Laboratory
frame and $x$ is known as the Bjorken variable. 
The DIS regime occurs in the so called
Bjorken limit:
\begin{equation}
 -q^2 = Q^2 \to \infty\quad , \quad \nu = E -
E^\prime \to \infty \quad , \quad x \, \, \rm {fixed}~.
\label{bj}
\end{equation}

The difference of the inclusive
cross-sections for
the lepton and the target nucleon polarized 
longitudinally, i.e. along or opposite to the
direction of the lepton beam,  is given by
\begin{equation} \label{eq:LongXsec}
\frac{d^2\sigma^{\begin{array}{c}\hspace*{-0.2cm}\to\vspace*{-0.2cm}\\
\hspace*{-0.2cm}\Leftarrow\end{array}}}{dx\,dy}-
\frac{d^2\sigma^{\begin{array}{c}\hspace*{-0.2cm}\to\vspace*{-0.2cm}\\
\hspace*{-0.2cm}\Rightarrow\end{array}}}{ dx\,dy}
 =
\frac{16\pi\alpha^2}{Q^2}\Biggl[\left(1-\frac{y}{2}-\frac{y^2 M^2 x^2}
{Q^2}\right)\,g_1(x,Q^2) - \frac{2M^2x^2y}{Q^2}g_2(x,Q^2)\Biggl] \,
\label{al}
\end{equation}
where the reversal of the nucleon's spin direction is indicated by 
the double arrow.
% We are neglecting, as is customary, the lepton masses.
The functions $g_1(x,Q^2)$ and $g_2(x,Q^2)$ are called 
spin dependent structure functions.
Another difference of cross sections, similar to Eq. (\ref{al}),
can be defined for transversely polarized nucleons,
%in the nucleon plane which allow measurement of both
yielding a different combination of $g_1(x,Q^2)$ and $g_2(x,Q^2)$.
In experiments with both longitudinal and
transverse target polarization, both $g_1 $ and $g_2$
can be therefore measured.
In Eq. (\ref{al}) it is anyway easily seen that the contribution
of $g_2(x,Q^2)$ is vanishing in the Bjorken limit, Eq. (\ref{bj}), 
and it is therefore difficult to measure
it in actual DIS experiments. For this reason, only in recent years 
it has been possible
to gather precise information on $g_2(x,Q^2)$. In this presentation, 
only the longitudinal
polarization structure function $g_1(x,Q^2)$ will be discussed.

\section{The parton model and QCD}

The parton model was proposed by Feynman (Feynman, 1972 \cite{sei}), 
a few years before QCD, to explain Bjorken scaling, i.e., the
fact that the structure functions in the Bjorken limit depend 
almost only on $x$ and not on $Q^2$.
The partons are pointlike constituents of the nucleon, 
which interact electromagnetically like
leptons. The  nucleon, in a frame where it is moving very fast, 
could be viewed as a ``beam" of collinear partons.
The partons are characterized as having momentum 
$p = x' P $, where $P$ is the momentum of the nucleon,
and covariant spin vector $s$. The interaction with the hard photon is 
then visualized as in Fig. 2,
in which the lepton-parton scattering is treated analogously 
to elastic lepton-lepton scattering.
$S$ is the covariant spin vector of the nucleon. 
Requiring the final parton to be on mass shell,
i.e. $(p+q)^2=0$, selects the value $x'=x$. 
Thus $x$ can be interpreted as the fraction of longitudinal
momentum of the target carried by the struck parton.
The partons were shown to have the quantum numbers of the current quarks, 
which were found to carry
only a part of the nucleon momentum. The missing part was
ascribed to a neutral vector particle, the gluon, which only carries 
color degrees of freedom, and
therefore does not couple to leptons.

\begin{figure}[htb!]
\centerline{\epsfig{file=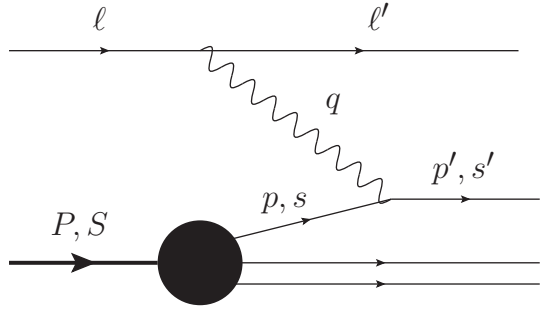,scale=0.5}}
\caption{Parton model description of DIS.}
\end{figure}

For unpolarized DIS, one finds the scaling result expressed in 
terms of the number density
of quarks, $q(x)$, and antiquarks,
$\bar q(x)$
\be
 F_1(x) = \frac{1}{2}\, \sum_j \, e^2_j \,[\, q_j(x)
+ \bar q_j(x)]\, , 
\ee 
where the sum is over flavors $j$,
$e_j$ is the charge of the quarks, and the Callan-Gross
relation
\be
 F_2(x)=2xF_1(x)\, ,
\ee
implies that the charged partons are spin $1/2$ particles.

For longitudinally polarized DIS one obtains
\be \label{eq:g1partonmodel}
 g_1(x) = \frac{1}{2}\, \sum_j \,
e^2_j\,[ \triangle q_j(x) + \triangle \bar q_j(x)]\, , 
\ee
where
\be
 \triangle q(x) = q_{\rightarrow}(x)- q_{\leftarrow}(x)\, , \ee
are the longitudinal polarization functions, or helicity dependent
parton distributions,
with $q_{\rightarrow(\leftarrow)}(x)$ representing the number densities 
of quarks whose spin
orientation is parallel (antiparallel) to the
longitudinal spin direction of
the proton (see Fig.~3). In terms of these, the
unpolarized parton density is
\be 
\label{eq:qvsqplusminus} q(x) = q_{\rightarrow}(x) + q_{\leftarrow}(x)\, . 
\ee

\begin{figure}[htb]
\centerline{ \epsfig{file=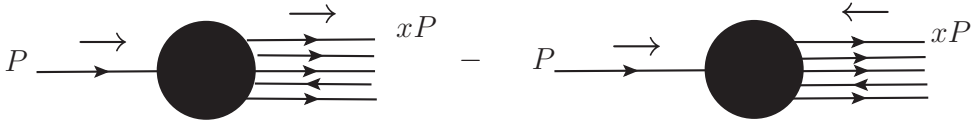,scale=0.8}}
\caption{Visualization of the longitudinally polarized parton
density $\Delta q(x)$. The upper arrows show the spin
direction.}
\end{figure}

The parton model is an intuitive description, much like the impulse
approximation in nuclear physics, and it appeared long before QCD.
Once QCD is accepted as the theory of strong interactions, with quark and
gluon as the fundamental fields, there  are interaction
dependent modifications of the simple parton model results for DIS.
%The theory is invariant under color gauge transformations,
%but the physical content of individual Feynman diagrams does
%depend upon the gauge, and it turns out that a parton-like picture
%emerges in the light-cone gauge $A_{+} =0$ where $A^\mu $ is the
%gluon vector potential.
The main impact of the QCD interactions is: i) to introduce
calculable logarithmic $Q^2$ dependence in the parton densities; ii)
to generate a contribution of the gluons to the structure functions,
and, in particular to $g_1$, arising from the polarization of
the gluons in the nucleon.

\begin{figure}[htb!]
\centerline{\epsfig{file=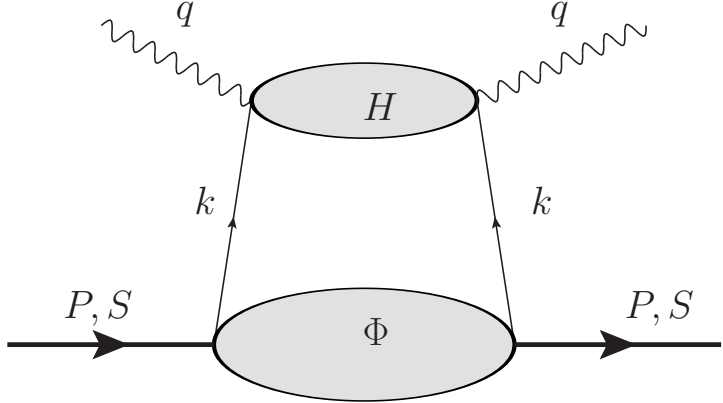,scale=0.6}}
\caption{QCD generalization of the parton model. 
In the factorized scheme, $H$ represents the hard part,
while $\Phi$ the soft part.}
\end{figure}

Unfortunately, these correction terms are infinite. The infinity is caused by 
collinear divergences which occur because of the masslessness of the quarks 
and which are removed through a mechanism called factorization of collinear 
divergences. This mechanism allows the reaction to be written as a product of 
a hard and a soft part (see Fig. 4), and the infinity is absorbed in 
the latter, which cannot be calculated in QCD. 
For this part either models of hadrons or parametrizations of data are used.
After an elaborate calculation, $g_1$ is found to depend also on 
$Q^2$:
\begin{eqnarray}
 g_1(x,Q^2) & = &  \frac{1}{2}\sum_{flavors}\, e_q^2 \, 
\Big \{  \Delta q(x,Q^2) +  \Delta \bar {q}(x,Q^2)  \nonumber \\
 & + & \frac{\alpha_s(Q^2)}{2 \pi}\int_x^1 \, \frac{dy}{y}\{ 
\Delta C_q(x/y) \, [\Delta q(y,Q^2) +  \Delta \bar {q}(y,Q^2) ]  \nonumber \\
 & + & \Delta C_G(x/y)\, \Delta G(y,Q^2) \} \Big \}
\end{eqnarray}
where  $\alpha_s(Q^2)$ is the QCD running coupling constant, while
$\Delta C_G$ and $\Delta C_q $ are the so called Wilson coefficients,
which can be evaluated in pQCD.
One can notice that the first line in the above equation represents 
the result obtained in QCD at leading order, while the remaining part 
is given by the higher-order contributions. Both the helicity dependent 
parton distributions and the Wilson coefficients depend on the factorization 
and renormalization schemes used.

\section{The proton spin problem}

Let us return to the parton model and write the expression of $g_1$,
for protons ($p$) and neutrons
($n$), respectively,
in terms of linear combinations of
the helicity dependent parton distributions
associated with the quark flavors $u, d, s$:
\begin{equation}\label{eq:g1SU3}
  g_1^{p(n)} (x) = \frac{1}{9}\left[ \pm \frac{3}{4}\Delta q_3(x) +
  \frac{1}{4}\Delta q_8(x) +\Delta  \Sigma(x) \right]\,
\end{equation}
where
\begin{equation}\label{eq:Deltaq3} \Delta q_3 =
(\Delta u + \Delta\overline{u}) - (\Delta d + \Delta\overline{d})~,
\end{equation}
\begin{equation}\label{eq:Deltaq8}
 \Delta q_8 = (\Delta u + \Delta\overline{u}) + (\Delta d + \Delta\overline{d}) -2(\Delta s + \Delta\overline{s})~,
\end{equation}
\begin{equation}\label{eq:Deltasigma}
 \Delta \Sigma = (\Delta u + \Delta\overline{u}) + (\Delta d + \Delta\overline{d}) + (\Delta s + \Delta\overline{s})\, .
\end{equation}
These quark densities transform respectively as the third component of
an isotopic spin triplet, the eighth component
of an $SU(3)_F$ octet, and a flavor singlet.

Taking the first moment of Eq. (\ref{eq:g1SU3}) yields
\begin{equation}\label{eq:Gamma1}
  \Gamma _1^{p(n)} \equiv \int_0^1 g_1(x) dx = \frac{1}{9}\left[
\pm \frac{3}{4}g_A^3 +
  \frac{1}{4}g_A^8 + g_A^0 \right]~,
\end{equation}
where
\begin{eqnarray} \label{eq:a3,8,0}
g_A^3&=&\int^1_0 dx~\Delta q_3(x)~, \nonumber \\
g_A^8&=& \int^1_0 dx~\Delta q_8(x)~, \nonumber \\
g_A^0&= & \int^1_0 dx~\Delta\Sigma (x) .
\end{eqnarray}

The  octet of currents associated to the distributions
Eqs. (\ref{eq:Deltaq3})--(\ref{eq:Deltasigma})
is precisely the one that controls
the weak decays of the
neutron and of the octet hyperons, which implies that the
values of $g_A^3$ and $g_A^8$ are known from
$\beta$-decay measurements:

\be \label{eq:a38} g_A^3 \equiv g_A = 1.2670 \pm 0.0035\, , \qquad  g_A^8=0.585 \pm
0.025\, . \ee
Hence, according to Eq. (\ref{eq:Gamma1}), a measurement of $\Gamma_1$ can be
considered as giving the value of the flavor singlet $g_A^0$.

In the parton model, two remarkable sum rules can be obtained.
The Bjorken sum rule (Bjorken, 1970 \cite{sette}), 
initially derived using only the parton model and isospin invariance 
(see Eq. (\ref{eq:Gamma1})), reads, to leading order in QCD:
\beq \label{eq:Bjsum}
\int_0^1 dx\,[g_1^p(x,Q^2) - g_1^n(x,Q^2)] = \frac{g_A}{6}\, .
\eeq
QCD corrections to this result have been calculated up to third order in 
the coupling constant (Larin {\it et al.}, 1997 \cite{otto}) leading to
\beq \label{Bjqcd} 
\int_0^1 dx\,[g_1^p(x,Q^2) - g_1^n(x,Q^2)] =
\frac{g_A}{6} \left\{ 1- \frac{\alpha_S}{\pi} - 3.583 
\left( \frac{\alpha_S}{\pi} \right )^2 - 20.215 
\left ( \frac{\alpha_S}{\pi} \right )^3 \right \},
\eeq
a result which has been confirmed experimentally.

Besides, in the parton model
\be
 g_A^0  = g_A^8 + 3(\Delta s + \Delta\overline{s}),
\label{ga0}
\ee
and, if one neglects the
contribution from the strange quarks in Eq. (\ref{ga0}),
i.e., from $\Delta s + \Delta \bar{s}$,
the Ellis and Jaffe sum rule (Ellis and Jaffe, 1974 \cite{nove}) is obtained :
\be
g_A^0 \simeq g_A^8 \simeq 0.6~.
\ee

In 1987, the EMC Collaboration performed a measurement
of $g_1^p$ and $\Gamma _1^p$ and,
using the known values of $g_A^3$ and $g_A^8$ in Eq. 
(\ref{eq:Gamma1}), obtained
\be  g_A^0 \simeq 0 , \ee
a value in contradiction with the Ellis-Jaffe sum rule.
This experimental result has other dramatic implications.
Consider the physical significance of $\Delta\Sigma(x)$. Since
$q_{\rightarrow (\leftarrow)} (x)$ count the number of quarks of 
momentum fraction $x$
with spin component $\pm \frac{1}{2}$ along the direction of
motion of the proton (say the $z$-direction), the total
contribution to $J_z$ coming from the spin of a given flavor
quark is
\begin{eqnarray}\label{eq:Szfromquark}
\langle S_z \rangle&=& \int^1_0
dx\Biggl\{\Biggl(\frac{1}{2}\Biggl) q_\rightarrow(x) +
\Biggl(-\frac{1}{2}\Biggl) q_\leftarrow(x) \Biggl\} \nonumber\\
&=&\frac{1}{2} \int^1_0 dx~\Delta q(x) \,.
\end{eqnarray}
It follows that
\begin{equation}\label{eq:a0Sz}
g_A^0 = 2\langle S^{quarks}_z \rangle~,
\end{equation}
where $ \langle S^{quarks}_z \rangle $ is the contribution to
$J_z$ from the spin of all quarks and antiquarks.
The connection between $g_A^0$ and 
$\langle S_z^{quarks} \rangle $, and the possible implications
of the result $\langle S_z^{quarks} \rangle$ 
smaller than 1/2, were discussed in a paper by Sehgal
\cite{Sehgal:1974rz}.

In a non-relativistic constituent model one would naively
expect all of the proton spin to be carried by the spin of its
quarks. In a  relativistic model one expects
$2\langle S^{quarks}_z \rangle \approx 0.6$, due to the loss of
normalization of the upper components, in agreement
with the Ellis-Jaffe sum rule but far from the EMC
result for $g_A^0$. This contradiction was labeled as the proton
``spin crisis'', and it was the beginning of an impressive experimental
and theoretical activity, which is still going on.
From the theoretical side,
a careful reanalysis both of non relativistic and relativistic models
has lead to results in closer agreement to the EMC value, as we will show.
Another argument has been the observation that,
even in the Bjorken limit, the gluonic version of the anomalous 
triangle diagram leads
to gluonic contribution to the first moment of $g_1$ (Altarelli and Ross, 1988
\cite{dieci}):
\begin{equation}\label{eq:g1gluon}
\Gamma^{gluons}_{1}(Q^2) = -\frac{1}{3}\,
\frac{\alpha_s(Q^2)}{2\pi}\,\Delta G(Q^2)\, .
\end{equation}
This result is of fundamental importance since it
implies that the simple parton model formula
for $g_A^0$ is incomplete. Instead, one has
that the quantity measured by the EMC collaboration is
\begin{equation}
\tilde{g}_A^0 = g_A^0 -3 \, \frac{\alpha_s}{2\pi}~\Delta G~,
\end{equation}
allowing for the EJ sum rule to be fulfilled for a large enough 
value of $\Delta G$.
However, this result depends on the factorization scheme utilized. 
In schemes where $g_A^0$ has the meaning of a spin, 
the small measured $\tilde{g}_A^0$
does not necessarily imply  that the physically meaningful 
$g_A^0$ is small.
This observation was initially presented as a possible 
resolution of the spin crisis,
but it is now clear that it is not sufficient. As a matter 
of fact, to explain the measured value of $g_A^0$, 
it should be $\Delta G \approx 1.7$, while the 
observed values are much too small ($|\Delta G| \approx 0.29$) 
to resolve the spin crisis, and different analyses give even 
different signs. A way out of this problem is to consider that the 
partons possess orbital angular momentum.

From the experimental side, after the publication of the EMC data, 
an impressive program started
in several laboratories to extend the kinematical range of the EMC 
experiment, to reduce
the systematic errors in the measurement of $\Gamma _1^p$, and to get
the neutron information, necessary to test the fundamental Bjorken Sum Rule,
Eq. (\ref{eq:Bjsum}).
In this way, it was possible to realize if the spin crisis could be ascribed
to a problem of the parton model and, in turn, of the underlying theory, QCD.
The neutron measurement is really difficult due to the lack of pure
neutron targets, so that nuclear targets have to be used and nuclear structure
effects have to be carefully taken into account. Despite of these difficulties,
a reasonable amount of precise data is nowadays available for the neutron, 
as it will be seen in the following.

\section{Models of hadron structure}

In this section, it will be shown that properly built models can
help in clarifying the origin of the so-called spin crisis.
QCD  is a theory of quarks  (antiquarks) and gluons,
as has been shown in the asymptotic regime,
where the interaction can be treated perturbatively. 
At low energies, the idea that
baryons are made up of three {\it constituent} quarks and mesons of a
{\it constituent} quark-antiquark pair, the naive quark model
scenario, accounts for a large number of experimental facts.
The quest for a relation between  the ''current'' quarks of the theory
and the {\it constituent} quarks of the model has an old history and
this search has been the {\it leitmotiv} of a considerable research effort.
The fundamental problem one would like to understand is how confinement,
i.e. the apparent absence of color charges and  dynamics in hadron physics,
is realized.

\begin{figure}[htb]
\begin{center}
\vspace{0.5cm}
%\epsfig{file=three.eps,scale=0.27,angle=270}
%\hskip 0.5cm
\epsfig{file=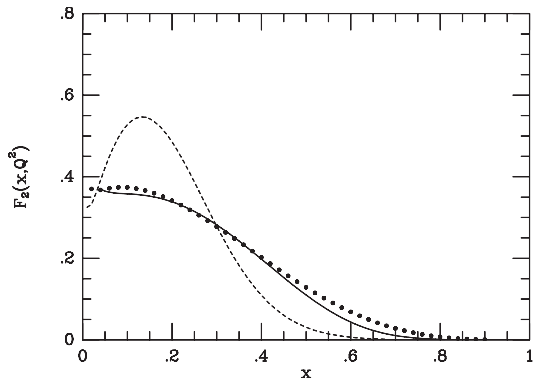,scale=0.27,angle=270}
\caption{
The structure function $F_2(x,Q^2)$ obtained
by NLO-evolution to $Q^2=10$ GeV$^2$ in a model with point-like quarks (dashed)
and in the convolution approach with composite quarks (full) using 
the wave functions
of a semirelativistic model. See 
%S. Scopetta, 
%V. Vento and M. Traini, Physics Letters, volume B 421, pages: 64-70, 1998
Ref \cite{io1}  
for details.\label{unpol}}
\end{center}
\end{figure}

Detailed quark models of hadron structure based on the constituent quark
concept have been defined in order to explain low energy properties.
In order to reach the high energies of the DIS regime,
the central assumption
is the existence of a scale,
$\mu_0^2$, where the short range (perturbative) part of the
interaction is negligible, and therefore the glue and sea are suppressed,
and the long range (confining) part of the interaction produces a
proton composed of three (valence) quarks only.  
One then ascribes the quark model
calculations of matrix elements to that hadronic scale
$\mu_0^2$. For larger $Q^2$ their Wilson coefficients will give the
evolution as dictated by pQCD. In this way quark models,
summarizing a great deal of hadronic properties, may substitute ''ad hoc''
low-energy parameterizations. The procedure describes successfully the gross
features of the DIS results.

In order to produce a more quantitative description of the data, 
different mechanisms
have been proposed: sea gluons,
sea quarks and antiquarks, meson clouds, relativistic effects, etc...
Some of these mechanisms appear naturally if one endows
the constituent quarks with structure. In this scenario, 
the constituent quarks are
themselves complex objects whose structure functions are described by a set of
functions $\Phi_{ab}$ that specify the number of point-like partons
of type $b$, which are present in the constituents of type $a$ with fraction
$x$ of its total momentum. In general $a$ and $b$
specify all the relevant quantum numbers of the partons, i.e., color, 
flavor and spin (see Altarelli {\it et al.}, 1974 \cite{11}, 
where this scenario has 
been firstly addressed. See also Hwa, 1980 \cite{12}).

\begin{figure}[htb]
\begin{center}
\vspace{0.5cm}
\epsfig{file=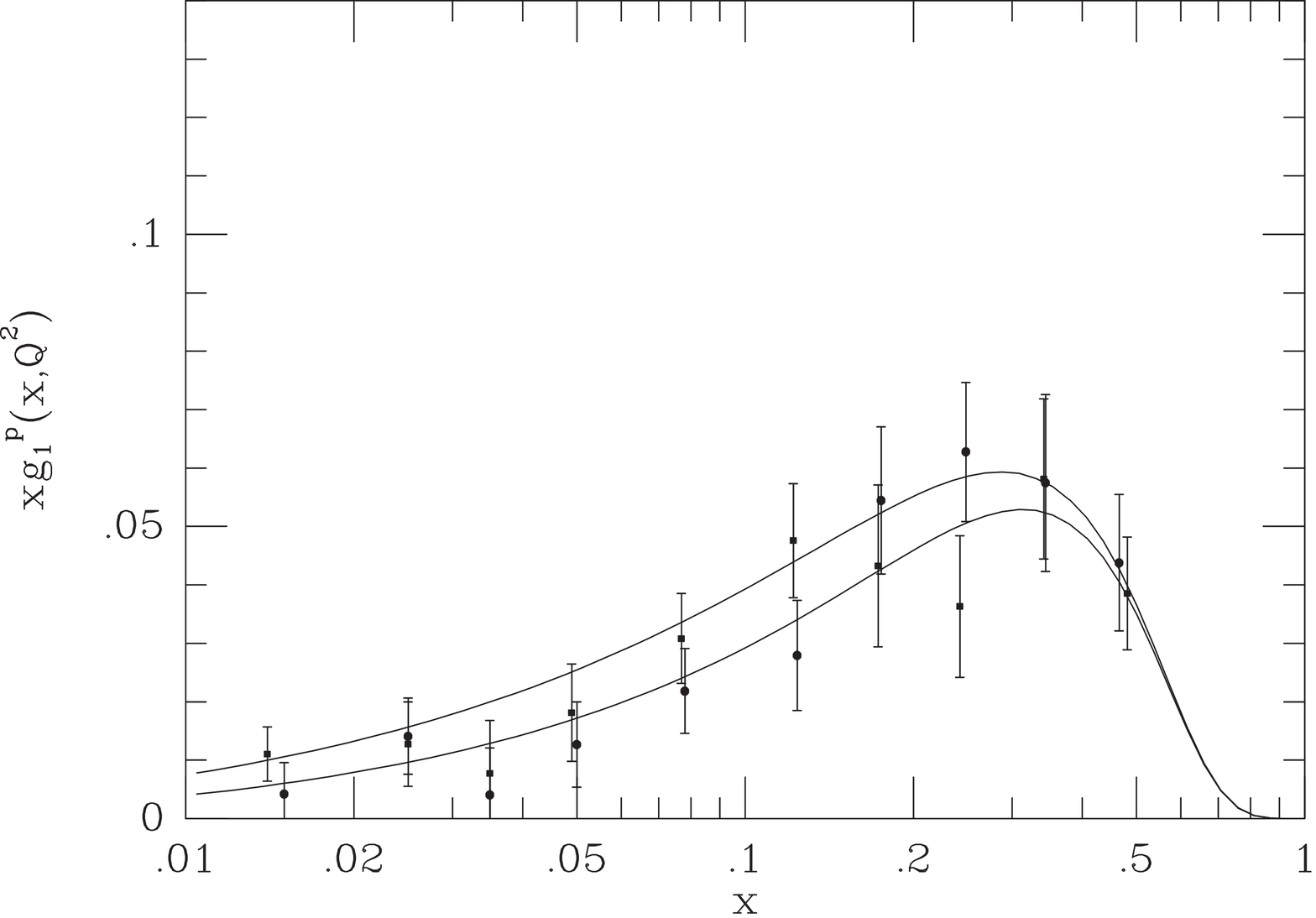,scale=0.27,angle=270}
\hskip 0.5cm
\epsfig{file=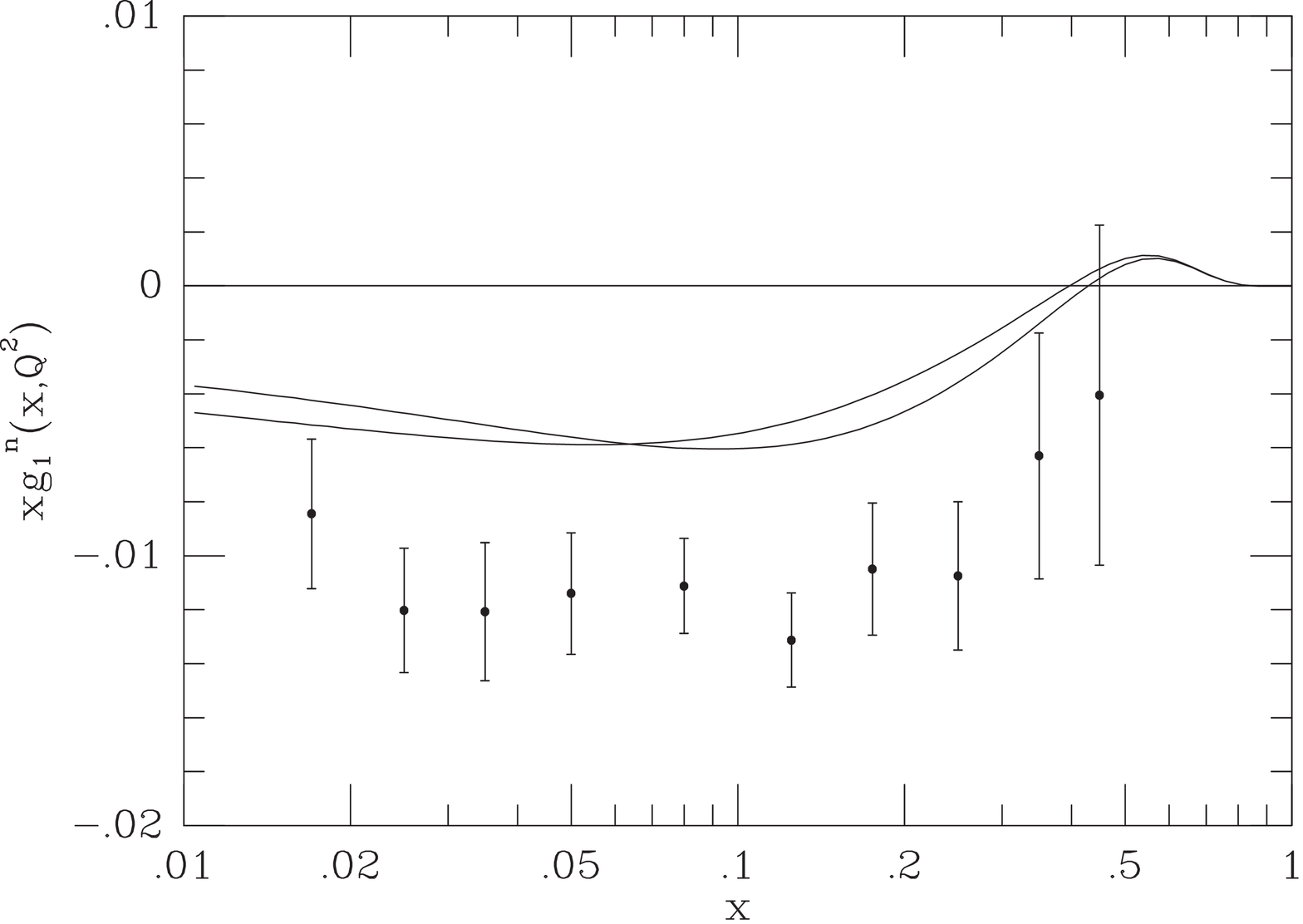,scale=0.27,angle=270}
\caption{
Left (Right): $xg_1(x,Q^2)$ for the proton
(neutron) evolved at NLO to
$Q^2= 10\,(5)$ GeV$^2$, for two different convolution models (full curves).
The proton (neutron) data are from the EMC and SMC collaborations
(from the E154 colaboration at SLAC). 
See 
%S. Scopetta, V. Vento and M. Traini, Physics Letters, 
%volume B 442,pages: 28-37, 1998 
Ref \cite{io2} for details.\label{polar}}
\end{center}
\end{figure}

Results are shown in Figs. \ref{unpol} and \ref{polar}.  
Fig. \ref{unpol} refers to the unpolarized case.
The structure function $F_2(x,Q^2)$, calculated
evolving the parton distributions obtained at low energy $u_o$ 
and $d_o$ within a semirelativistic model,
describes successfully the data. Moreover, the agreement becomes impressive
when compared with the analogous calculation with non-composite constituents.
A qualitatively similar agreement is obtained also
in the polarized case,
as it is shown in Figs. 6 and 7.
It should be noticed that in this framework
the {\it spin crisis}, as initially presented, does not arise:  the
{\it constituent} quarks, the effective degrees of freedom of the model,
can carry most of the proton spin, even if the measured $g_A^0$, a 
{\it current} quark DIS observable,
is small.
In this scenario, there is no need of large orbital angular momentum 
contributions to explain the data, once the proper structure of the 
constituent quarks has been taken into account.

In general, the non relativistic constituent quark models describe 
the nucleon by S-wave functions, or with small D-wave admixtures, 
once the hyperfine interaction is taken into account. Relativistic 
constituent quark  models have a non-zero quark orbital angular momentum 
from the beginning and reduce the valence quark spin contributions to 
the nucleon spin from 1 to about 0.6. However, to reach the scale of the 
data, perturbative QCD evolution has to be used and the contribution of 
quarks and gluons to the orbital angular momentum may be of different size, 
and relevant at the experimental scale. However, the perturbatively 
generated gluon orbital angular momentum is cancelled by the gluon 
helicity contribution, as it can be shown by general arguments.

\section{Experimental determination of the longitudinal
polarization functions}

\begin{figure}[htb]
\begin{center}
\epsfig{file=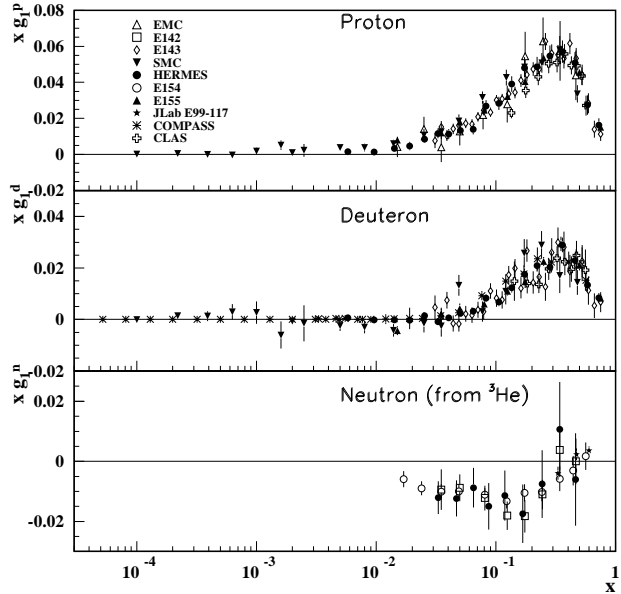,scale=0.45}
\caption{World data on the polarized structure function $x g_1(x)$
for the proton, deuteron and neutron
in the DIS region ($W > 2$~GeV), taken by
different experiments, at several different values of $Q^2$, as compiled 
in Ref. \cite{pdg}.
\label{g1world}}
\end{center}
\end{figure}

A vast amount of data on the inclusive spin structure function $g_1(x,Q^2)$
has been accumulated by lepton scattering experiments at SLAC, CERN, DESY
and Jefferson Lab (see Fig. 7).

\begin{figure}[htb]
\begin{center}
\epsfig{file=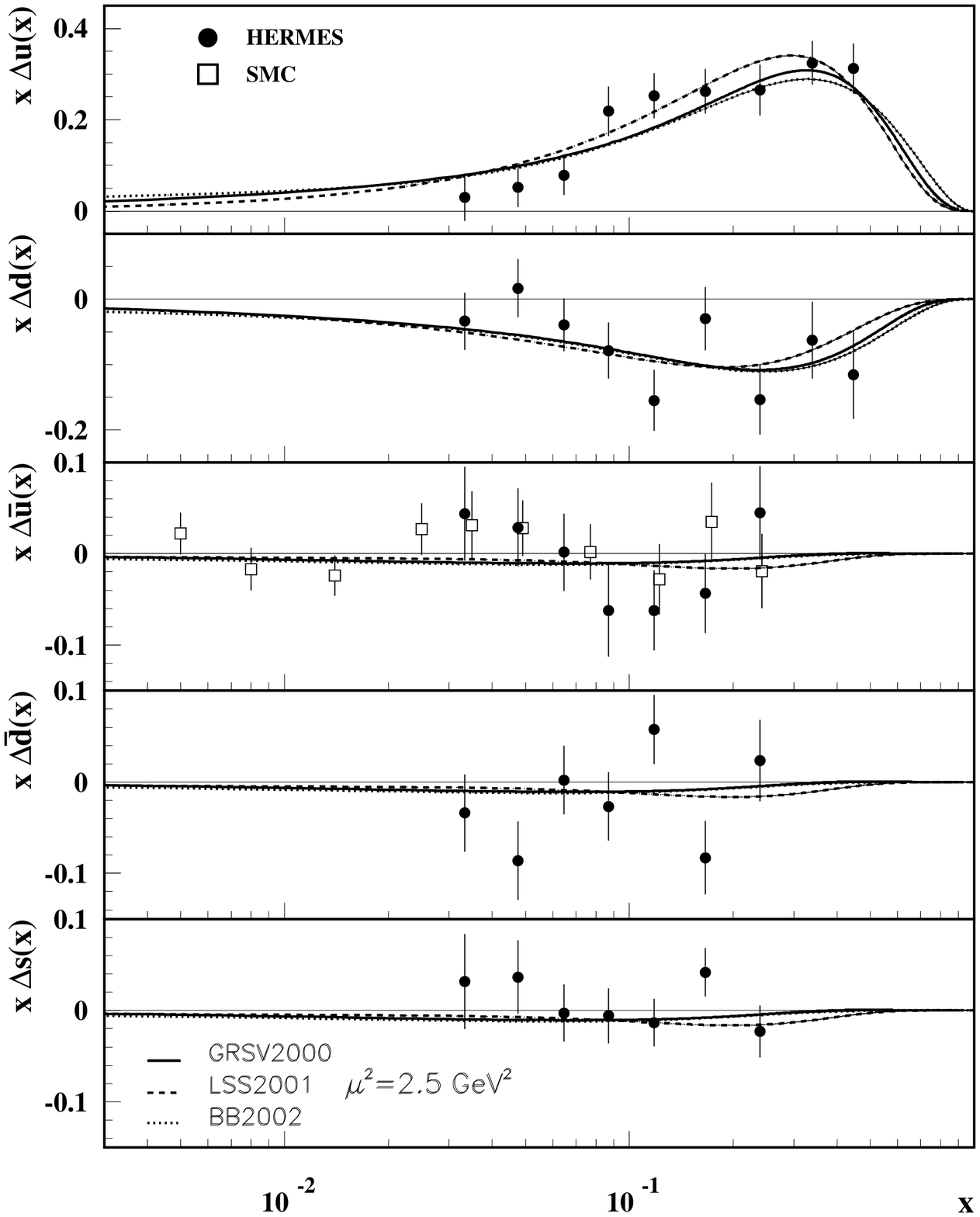,scale=0.45}
\caption{World data on helicity dependent parton distributions $\Delta q(x)$
extracted from semi-inclusive DIS data, as compiled 
in  Ref. \cite{pdg}.
\label{SIDISworld}}
\end{center}
\end{figure}

Inclusive data from proton, neutron, deuteron and perturbative QCD 
analysis  are used to extract the contributions from quark, antiquark and 
gluon densities with a precision that depends very strongly on that of the 
data. One approach to gather additional information has been to use 
semi-inclusive lepton scattering (SIDIS), where, in addition to the 
scattered lepton, one detects a leading hadron (typically a pion or kaon) 
in the final state. This approach was pioneered by the SMC collaboration 
and the most detailed data set stems from the HERMES and COMPASS 
collaborations (see Fig. 8). However, with SIDIS alone it is difficult to 
extract the sea and glue polarization. An alternate route has been employed 
in the polarized proton collision program at BNL, using the Relativistic 
Heavy Ion Collider (RHIC). One selects final state signatures like hadrons, 
jets or even direct photons with large transverse momentum that indicate 
an underlying hard interaction between two constituents from the two 
colliding protons. Recent analyses take into account all the data from 
inclusive, semi-inclusive polarized deep inelastic scattering and polarized 
pp scattering at RHIC, and perform all theoretical calculations at 
next-to-leading order of perturbative QCD to maximally constrain 
the extracted distributions (De Florian {\it et al.}, \cite{13}), 
opening the door to 
obtaining a better and more reliable picture of the spin structure 
of the nucleon.

\section{Summary}

After 30 years of dedicated experiments,
a rather detailed
picture of the nucleon spin structure is arising.
It appears that quark helicities do contribute a significant fraction
of the (longitudinal) spin of the nucleon. Most importantly, the fundamental
Bjorken sum rule and its pQCD evolution
are consistent with the data. None of the existing
experiments show any features that would contradict perturbative QCD 
in its realm of
applicability.

Most pQCD-based analyses of the data agree fairly well on the 
contribution of various
quark flavors to the proton spin, although there is still some 
controversy on the
role played by strange sea quarks. We also do not know yet whether
polarized up and down anti-quark densities show the same difference 
as the unpolarized ones.
The emergent picture is one where valence quarks
($\Delta q_V\equiv \Delta q - \Delta \bar{q} $) carry 
roughly the expected fraction $(\simeq 60\%)$ of the
nucleon spin, while the (on average) negative helicity of sea quarks reduces
this to about 30-35 %. Experiments
at Jefferson Lab
will extend our knowledge of polarized quark densities
out to $x>0.8$ and will decisively test predictions from pQCD 
and QCD-inspired models.
Additional information on individual quark and antiquark flavor contributions
to the nucleon spin will come from future experiments at RHIC (in particular
from direct $W^{\pm}$ production at 500 GeV center-of-mass 
energy) and
the new FAIR facility in Darmstadt, Germany as well
as semi-inclusive measurements with COMPASS and at Jefferson Lab.

In addition to logarithmic violations of scaling, expected from pQCD, the data
show some evidence for the so called higher twist contributions,
i.e., non-scaling contributions to the structure
function $g_1$, at intermediate $Q^2$ values.
Future measurements at Jefferson Lab
will  improve our knowledge of higher-twist matrix elements, and of
the second spin structure function,
$g_2$, and its moments.

At the lower end of the $Q^2$ scale, at the real
photon point,
the Gerasimov-Drell-Hearn (GDH) sum rule (Gerasimov, 1966
\cite{14}; Drell and Hearn, 1966 \cite{15})
has been well confirmed. The transition from this point to DIS,
through the region of intermediate $Q^2$, is under 
investigation, in particular at JLab.

One of the most important remaining open questions is where is the nucleon spin
that is not carried by quark helicities. We do not know the precise magnitude
and shape of the gluon contribution, and it could still be an important 
fraction of the total.
Improved statistics from the direct measurements
at RHIC and COMPASS will help dramatically.

The contribution from quark angular momentum is also an important
ingredient in the total spin balance of the nucleon. While a direct measurement
is not available, one can learn much about the transverse distribution 
and motion
of quarks from semi-inclusive measurements of single spin asymmetries. This
is a fairly new field, related to the transversity observable,
with a very rich potential and a rapidly growing body of
experimental data, but lies outside the scope of this article.

Another way to describe the nucleon spin is  the study of
Generalized Parton Distributions (GPDs), in particular
Deeply Virtual Compton Scattering (DVCS). Moments of certain combinations
of GPDs can be related to the total angular momentum (spin and orbital) carried
by various quark flavors, as expressed in Ji's sum rule (Ji, 1997 \cite{16}).
Data taken at Jefferson Lab are being analyzed and future experiments 
at COMPASS
will clarify the situation enormously.

In summary, the description of the nucleon spin keeps being a 
fundamental problem of hadron
structure (see Refs. \cite{17,18,19,20,21,22,23} 
and the links in Ref \cite{24} for further reading). 
An extensive and rich experimental program has been 
undertaken at existing facilities like
CERN (COMPASS), RHIC and Jefferson Lab and much theoretical effort is being 
devoted to a full
understanding of the nucleon spin structure.


\begin{thebibliography}{99}

\bibitem{uno}
J.D. Bjorken and E.A. Paschos, ``Inelastic Electron-Proton and 
$\gamma$-proton Scattering and the Structure of the Nucleon'', 
Phys. Rev. 185 1975, 1969. 

\bibitem{due} 
M. Breidenbach, J.I. Friedman, H.W. Kendall, (MIT, LNS), 
E.D. Bloom, D.H. Coward, H.C. DeStaebler, J. Drees, L.W. Mo, 
R.E. Taylor,`` Observed Behavior of Highly Inelastic electron-Proton 
Scattering", Phys. Rev. Lett. 23 935, 1969.

\bibitem{tre} 
H. Fritzsch, M. Gell-Mann and H. Leutwyler, ``Advantages of the Color 
Octet Gluon Picture'', Phys. Lett. B47 365, 1973.

\bibitem{quattro}
H.D. Politzer, ``Asymptotic freedom: an approach to strong interactions'', 
Phys. Rep. 14 129, 1974.

\bibitem{cinque} 
J. Ashman, {\it et al.}, [European Muon Collaboration],
``A Measurement of the Spin Asymmetry and Determination of the Structure 
Function g(1) in Deep Inelastic Muon-Proton Scattering'', 
Phys. Lett. B206 364, 1988.

\bibitem{sei}  R.P. Feynman,``Photon Hadron Interactions'' 
(Benjamin, New York, 1972).

\bibitem{sette}
J.~D.~Bjorken, ``Asymptotic Sum Rules At Infinite Momentum'', 
Phys. Rev. 179:1547, 1969 .

\bibitem{otto}
S.A. Larin, T. van Rittenberg and J.A.M. Vermasseren, 
``The Alfa-s**3 approximation of quantum chromodynamics to 
the Ellis-Jaffe sum rule'', Phys. Lett B404 153, 1997.

\bibitem{nove}  
J.R. Ellis, R.L. Jaffe,``A Sum Rule for Deep Inelastic 
Electroproduction from Polarized Protons'', Phys. Rev. D9 1444, 1974.

%\cite{Sehgal:1974rz}
\bibitem{Sehgal:1974rz}
L.~M.~Sehgal,
``Angular Momentum Composition of the Proton in the Quark Parton Model,''
Phys.\ Rev.\  D10, 1663, 1974.

\bibitem{dieci} G. Altarelli, G.G. Ross, ``The Anomalous Gluon Contribution 
to Polarized Leptoproduction",  Phys. Lett. B212 391, 1988.

\bibitem{io1}
S. Scopetta, V. Vento and M. Traini, 
``Towards a unified picture of constituent and current quarks'',
Phys. Lett. B421 64, 1998.  

\bibitem{11} G. Altarelli, N. Cabibbo, L. Maiani, R. Petronzio, 
``The nucleon as a bound 
state of three quarks and deep inelastic phenomena'', Nucl. Phys. B69  
531, 1974. 

\bibitem{12}
R.C. Hwa, ``Evidence for valence-quark clusters in nucleon 
structure functions'', Phys. Rev. D22 759, 1980.

\bibitem{io2}
S. Scopetta, V. Vento and M. Traini, 
``Polarized structure functions in a constituent quark scenario'',
Phys. Lett. B442 28, 1998.

\bibitem{pdg}
C. Amsler {\it et al.},
Particle Data Group, 
Phys. Lett. B667 1, 2008. 

\bibitem{13}
D. de Florian, R. Sassot, M. Stratmann and W. Vogelsang, 
``Global Analysis of Helicity Parton Densities and Their Uncertainties'', 
Phys. Rev. Lett. 101 072001, 2008. 

\bibitem{14}  S.B. Gerasimov,
``A Sum rule for magnetic moments and the damping of the nucleon 
magnetic moment in nuclei", Sov. J. Nucl. Phys. 2 430, 1966.

\bibitem{15}
S. D. Drell, A. C. Hearn, ``Exact Sum Rule for Nucleon Magnetic Moments", 
Phys. Rev. Lett. 16 908, 1966.

\bibitem{16} X.~D.~Ji, ``Gauge invariant decomposition of nucleon spin", 
Phys. Rev.  Lett.  78 610, 1997.

\bibitem{17} 
M. Anselmino, A. Efremov, E. Leader, 
``The theory and phenomenology of polarized deep inelastic scattering'', 
Physics Reports 261 1, 1995.

\bibitem{18} 
B. Lampe, E. Reya, 
``Spin physics and polarized structure functions'', 
Physics Reports 332 1, 2000.

\bibitem{19} 
B.W. Filippone and X. Ji,
``The spin structure of the nucleon'', 
Advances in Nuclear Physics 26 1, 2001.
 
\bibitem{20} 
S.D. Bass, 
``The spin structure of the proton'', 
Reviews of Modern Physics 77 1257, 2005.

\bibitem{21} 
B. Foster, A.D. Martin, M.G. Vincter, 
``Review of structure functions'', 
Review of Particle Properties 2009,  chapter 16.

\bibitem{22} 
S.E. Kuhn, J.-P. Chen, E. Leader, 
``Spin structure of the nucleon -- status and recent results'', 
Progress in Particle and Nuclear Physics 63 1, 
2009.

\bibitem{23} 
F. Myhrer and A.W. Thomas, 
``Understanding the proton's spin structure'', 
Journal of Physics G. Nuclear and Particle Physics 37 023101, 2010.

\bibitem{24}
Particle data group, http://pdg.lbl.gov/;    
Spin Muon Collaboration (SMC) at CERN, http://na47sun05.cern.ch/;    
Hermes at DESY, http://www-hermes.desy.de/;    
Compass at CERN, http://wwwcompass.cern.ch/;    
RHIC at BNL, http://www.bnl.gov/rhic/collaborators.asp;    
Hall A at TJNAF, http://hallaweb.jlab.org/physics/;    
CLAS at TJNAF,  http://www.jlab.org/Hall-B/;    
FAIR at GSI, http://www.gsi.de/fair/.    

 
\end{thebibliography}
\end{document}